\documentclass{aip-cp}

\usepackage[numbers]{natbib}
\usepackage{rotating}
\usepackage{graphicx}
\usepackage{mathrsfs}
\usepackage{amssymb}
\usepackage{graphicx}
\usepackage[normalem]{ulem}
\usepackage[dvips]{color}
\usepackage{bm}
\usepackage{longtable}
\usepackage{times}

\newcommand{\apj}{ApJ}
\newcommand{\apjl}{ApJL}
\newcommand{\mnras}{MNRAS}
\newcommand{\pasa}{PASA}
\newcommand{\pasj}{PASJ}
\newcommand{\nat}{Nature}
\newcommand{\aap}{A\&A}

\renewcommand\sout{\bgroup \color{red} \ULdepth=-.5ex \ULset}

\renewcommand{\rm}[1]{\textrm{#1}}

\begin{document}

% Title portion
\title{Studying newborn neutron stars by the transient emission after stellar collapses and compact binary mergers}
%: Super-luminous supernovae, gamma-ray bursts, and fast blue optical transients}
\author[aff1]{Yun-Wei Yu\corref{cor1}}
\corresp[cor1]{Corresponding author and speaker: yuyw@mail.ccnu.edu.cn}
\author[aff1]{Aming Chen}
\author[aff2]{Zi-Gao Dai}
\author[aff1]{Shao-Ze Li}
\author[aff2]{Liang-Duan Liu}
%\author[aff3]{Bing Zhang}
\author[aff3]{Jin-Ping Zhu}
%\eaddress{bjcai87@gmail.com}; \eaddress{Lwchen@sjtu.edu.cn}
\affil[aff1]{Institute of Astrophysics, Central China Normal
University, Wuhan 430079, China}
\affil[aff2]{School of Astronomy and Space Science, Nanjing University, Nanjing 210093, China}
%\affil[aff3]{Department of Physics and Astronomy, University of Nevada Las Vegas, Las Vegas, NV 89154, USA}
\affil[aff3]{Kavli Institute for Astronomy and Astrophysics and Department of Astronomy, Peking University, Beijing 100871, China}
\maketitle

\begin{abstract}
The formation of neutron stars (NSs), both from collapses of massive stars and mergers of compact objects, can be usually indicated by bright transients emitted from explosively-ejected material. In particular, if the newborn NSs can rotate at a millisecond period and have a sufficiently high magnetic field, then the spin-down of the NSs would provide a remarkable amount of energy to the emitting material. As a result, super-luminous supernovae could be produced in the massive stellar collapse cases, while some unusual fast evolving and luminous optical transients could arise from the cases of NS mergers and accretion-induced collapses of white dwarfs. In all cases, if the dipolar magnetic fields of the newborn NSs can be amplified to be as high as $10^{15}$ G, a relativistic jet could be launched and then a gamma-ray burst can be produced as the jet successfully breaks out from the surrounding nearly-isotropic ejected material.
\end{abstract}

% Head 1
\section{Neutron Stars and Transient Phenomena}\
From the pioneering work by Baade \& Zwicky (1934), it has been widely considered that neutron stars (NSs) can be born from the core collapse of massive stars, accompanied by a luminous supernova emission. On the one hand, so far, thousands of NSs have been identified from our Galaxy through observations of pulsars, X-ray sources, etc. However, these Galactic NSs are generally older than a few thousand years. On the other hand, we can in principle explore newborn NSs by observing supernova emission and, meanwhile, thousands of supernovae have been detected from other galaxies beyond the Milky Way. Here, the problem is that the newborn NSs are always hid in a thick and dense supernova ejecta and usually have no impact on the supernova emission. The supernova emission is generally determined by the synthesization of nickels, the recombination of hydrogen and helium, and the interaction of the supernova ejecta with circum stellar medium (CSM). Therefore, in any case, our knowledge of NSs at their birth is actually very poor.

Fortunately, this difficult situation might have being changed by the discovery of
some unusual transient phenomena that are driven by the spin-down of a rapidly rotating and usually highly-magnetized NS. Such transient phenomena include super-luminous supernovae (SLSNe), gamma-ray bursts (GRBs), and mysterious fast evolving and luminous optical transients (FLTs). Therefore, it is definitely interesting and necessary to uncover how the newborn NSs can influence these transient emission and, simultaneously, to dig into the nature of these NSs by analyzing the transient observations.

\section{Super-luminous Supernovae}\label{sec2}
SLSNe are an unusual type of supernovae about $10-100$ times brighter than the normal ones (Gal-Yam et al. 2012). If the total
radiated energy of a typical SLSN of $10^{51}$
erg is provided by radioactive decays of $^{56}$Ni as usual, then the mass of the nickels should be about several to several tens of solar masses, which is nearly impossible for normal supernova
nucleosynthesis (e.g. Umeda \& Nomoto 2008). In principle, for some extremely massive stars, such a high mass of $^{56}$Ni could be still produced by their peculiar
supernova explosions triggered by electron-positron
pair-production instability (Barkat et al. 1967; Heger \& Woosley
2002). Nevertheless, in comparison with observational data, this pair-instability supernova model is usually disfavored by the rapid increase of SLSN light curves (Nicholl et
al. 2013) and the relatively low redshifts (McCrum et
al. 2014). Therefore, non-radioactive energy sources are definitely needed to account for the most observed SLSNe.

On the one hand, the narrow lines existing in the spectra of some hydrogen-rich SLSNe imply that these SLSNe must be primarily or, at least, partially powered by the shock interaction between a supernova ejecta and a dense CSM (Smith \& McCray 2007;
Moriya et al. 2011, 2013; Chevalier \& Irwin 2011; Ginzburg \&
Balberg 2012). The broad-lined features in the spectra further indicate that the velocities of the ejecta and thus their kinetic energies can indeed be high enough to power the SLSN emission. On the other hand, a long-lived
central engine has been widely suggested to be responsible for the most observed SLSNe, in particular, for the hydrogen-poor ones. Qualitatively, the central engine could be a spinning-down NS (Ostriker \& Gunn 1971; Woosley
et al. 2010; Kasen et al. 2010, 2016; Moriya et al. 2016; Chen et
al. 2016b) or a fallback accretion disk (Dexter \&
Kasen 2013). However, by fitting the SLSN light curves quantitatively, it can still be found that the NS model has an obvious advantage over the fallback accretion model, because of the wide ranges of parameters allowed by the NS model (Yu \& Li 2017). Therefore, in literature, the NS engine model has been frequently employed to account for SLSN observations, which nearly always went to a great success
(e.g., Dessart et al. 2012; Nicholl et al. 2013; Howell et al. 2013;
McCrum et al. 2014; Dai et al. 2016; Inserra et
al. 2013, 2016b; Chatzopoulos et al. 2013; Wang et al. 2015; Nicholl
et al. 2015a; Yu et al. 2017).

\begin{figure}
\centering\resizebox{0.7\hsize}{!}{\includegraphics{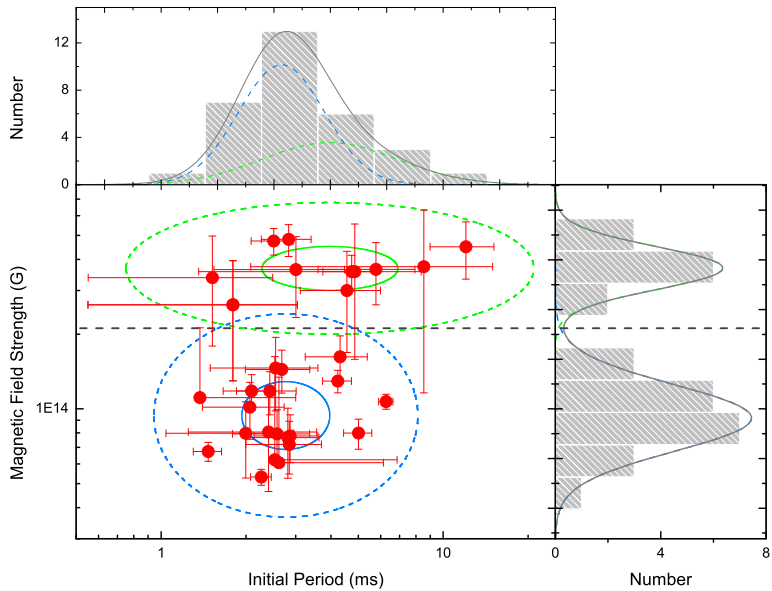}}
\caption{The magnetic filed strengths vs the
initial spin periods of NSs forming from SLSNe (Yu et al. 2017).}\label{fig1a}
\end{figure}

Due to their unusual energy supplies, the shape of the light curves of SLSNe would be mainly coined by the spin-down of the central NSs rather than by the radioactive processes as usual. This makes it feasible to find out the properties of the newborn NSs (e.g., the spin-down luminosities and timescales) from the observed SLSN light curves. Figure \ref{fig1a} presents the magnetic field strengths and the initial spin periods of NSs derived from 31 SLSNe. As shown, these NSs are all millisecond magnetars. To be specific, the magnetic field strengths are generally higher than the critical field strength $B_{\rm c}$ of electron
Landau quantization, while the spin periods range from 1 ms to 10 ms. These parameter values make these newborn NSs completely different from those discovered in the Galaxy. Moreover, by according to a cluster analysis on the data, we further found that the SLSN NSs could fall into two subclasses by an empirical separating line at $\sim5B_{\rm
c}$. This clustering can help to understand a plausible classification of SLSNe into
slow evolving and fast evolving ones, because the widths of the light curves of NS-powered SLSNe are primarily determined by the magnetic fields of the NSs (i.e., $\Delta t_{10\%}\propto B_{\rm p}^{-0.68}$). Following this consideration, it can be argued that the slow and fast evolving classification of SLSNe does not indicate different origins for them. In addition, we found that the rotational energies of the NSs forming from SLSNe are correlated with the masses of the supernova ejecta. This correlation gives a clue to explore the explosion mechanisms of these unusual supernovae.

%\begin{figure}
%\centering\resizebox{0.5\hsize}{!}{\includegraphics{fig1b1.eps}}\resizebox{0.5\hsize}{!}{\includegraphics{fig1b2.eps}}
%\caption{A collection of fast evolving (left) and slow-evolving
%(right) SLSN light curves (Yu et al. 2017).}\label{fig1b}
%\end{figure}

\begin{figure}
\resizebox{\hsize}{!}{\includegraphics{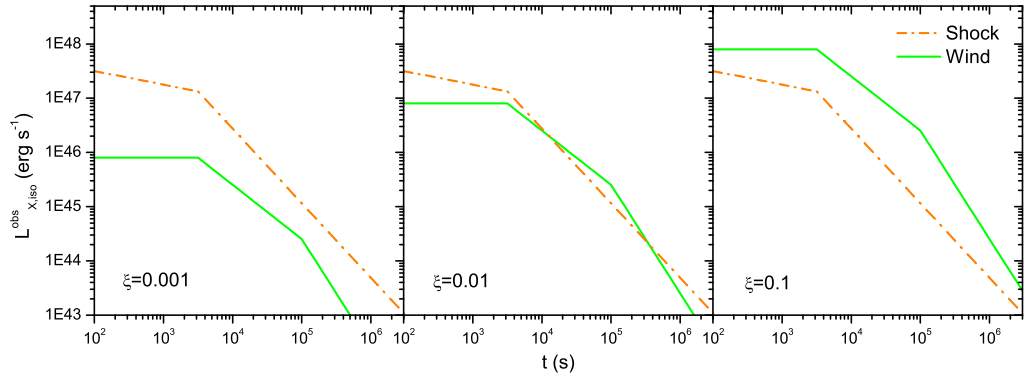}}
\caption{GRB afterglow light curves contributed by the PW emission and the external shock (Yu et al. 2010).}\label{fig2b}
\end{figure}

\begin{figure}
\centering\resizebox{0.5\hsize}{!}{\includegraphics{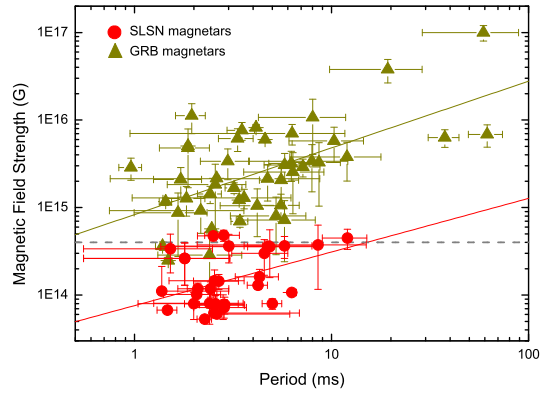}}
\caption{A comparison between the NS parameters of SLSNe and
LGRBs (Yu et al. 2017).} \label{fig2c}
\end{figure}

\section{Long-duration Gamma-ray Bursts}
After the above discussion on SLSNe, we would very like to mention that long-duration gamma-ray bursts (LGRBs) also belong to an drastic explosion phenomenon due to the core collapse of massive progenitors, which has been confirmed by their associations with broad-lined supernovae. Meanwhile, a remnant millisecond magnetar was also usually invoked to explain the shallow-decay and plateau afterglows of a remarkable number of LGRBs, by ascribing these emission component to a continuous energy release from the NS (Dai \& Lu 1998a, b; Zhang \& Meszaros 2001; Fan \& Xu
2006; Yu \& Dai 2007; De Pasquale et al. 2007; Metzger et
al. 2007; Troja et al. 2007; Lyons et al. 2009; Yu et al. 2010). In addition, the existence of post-GRB NSs has further been strongly supported by the observed sharp flares in GRB afterglows. These flares indicate that the GRB engines must have long-delayed intermittent activities. The NS nature of the engines is obviously reasonable and acceptable for explaining these activities (Dai et al. 2006). It is interesting and necessary to reveal the possible connections and
differences between the NS engines of SLSNe and LGRBs, in particular, in view of their similar collapse origins.

%\begin{figure}
%\resizebox{0.7\hsize}{!}{\includegraphics{fig2a.eps}} \caption{A schematic cartoon of an RWB. The meaning of
%the four regions (1-4) is explained in the text. In the inner (a) part of region 4, the energy outflow is
%dominated by low-frequency EM waves, in the outer (b) part of region 4 by a kinetic-energy flow carried by
%ultrarelativistic electron-positron pairs.}
%\end{figure}

While the afterglows of GRBs are usually ascribed to the emission of an external shock driven by the GRB jets, the plateau X-ray afterglow followed by a very steep decay, which was first discovered after GRB 070110, indicates that this plateau afterglow should have an internal origin, somewhat in analogy to the afterglow flares. To be specific, in the NS engine model, such internal-origin afterglows are probably produced by a relativistic wind from the NS, e.g., through magnetic reconnections in the pulsar wind (PW; Drenkhahn 2002, Drenkhahn \&
Spruit 2002) and/or a termination shock due to the collision of the PW with the GRB jet (Dai 2004, Yu \& Dai 2007). The PW emission can penetrate through and escape from the GRB jet freely, because the GRB jet is always transparent. This is very different from the supernova situations, where supernova ejecta are very opaque at early times. Following this consideration, Yu et al. (2010) suggested that, as long as a NS is formed, the subsequent afterglow emission should always be contributed by both internal and
external emission processes, not only for GRB 070110. And just the competition between the internal and external components determines the diversity of the
observed afterglow light curves, as illustrated by Figure \ref{fig2b}. In any case, no matter the afterglow emission is shock-dominated or PW-dominated, the properties of the NS engines can always be derived by fitting the afterglow light curves.
%the shock-dominated and wind-dominated emissions can lead to the one-break and two-break afterglow light curves, respectively.

As a result, the magnetic
fields of the NSs forming from LGRBs are found to be roughly within the range of $\sim (10-300)B_{\rm c}$ (L\"u \& Zhang 2014), which are generally much higher than those of SLSN NSs, as shown in Figure \ref{fig2c}\footnote{It worths to be mentioned that, for an ultra-high magnetic field leading to a significant stellar deformation, the spin-down of the NS can be dominated by gravitational wave (GW) radiation rather than by the usually-considered magnetic dipole radiation (MDR). In this case, the NS parameters presented in Figure \ref{fig2c} need to be revised.}. This strongly hints that a high magnetic filed is indispensable for the formation of a relativistic GRB jet. Meanwhile, because of the ultra-high magnetic fields, the spin-down timescales of LGRB NSs are much shorter than the thermal diffusion timescale of the isotropic supernova ejecta and thus the spin-down energy would be mainly converted into the kinetic energy of the supernova ejecta. Consequently, the observed LGRB-associated supernovae usually have broad-lined features but do not have a luminosity comparable to those of SLSNe. Additionally, for an extremely high magnetic field, the number of electron Landau
levels will become only several tens and even a few, which means the electrons in the NS interior would behavior very close to a
one-dimension gas. In this case, the resultant supernova explosion could be expected to be highly
anisotropic.

\begin{figure}
\centering\resizebox{0.9\hsize}{!}{\includegraphics{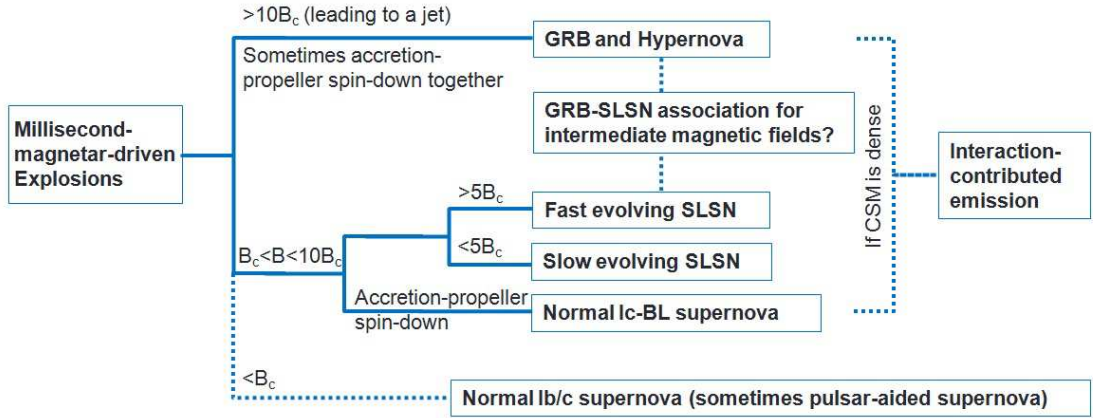}}
\caption{Possible connections between different magnetar-driven
explosion phenomena (Yu et al. 2017).}\label{fig2d}
\end{figure}

Finally, by according to the parameter distributions of the NSs of SLSNe and LGRBs, Yu et al. (2017) proposed a united picture to bring SLSNe and LGRBs into a common origin, as illustrated in Figure \ref{fig2d}. The differences between these different explosion phenomena are suggested to primarily arise from the different magnetic fields of their remnant NSs.

\section{Short-duration Gamma-ray Bursts and Mergernovae}\label{sec4}

Different from LGRBs, short-duration gamma-ray bursts (SGRBs) have been long hypothesized to originate from the mergers of double NSs or a NS and a black hole. Recently, the GW detection for GRB 170817A by LIGO has clearly confirmed that the progenitor of this SGRB is a NS-NS binary. Although the products of double NS mergers are usually suggested to be a black hole because of its high mass, we would still like to point out that the shallow-decay, plateau, and flare emission components actually widely exist in the afterglows of a remarkable number of SGRBs. Therefore, as discussed in the previous section, it is not unreasonable to suspect that the remnant objects of SGRBs can be a rapidly rotating, highly-magnetized, and massive NS (Dai et al.
2006; Fan \& Xu 2006; Rowlinson et al. 2010,
2013; Bucciantini et al. 2012; Gompertz et al. 2013, 2015;
Zhang 2013; L{\"u} et al. 2015). This supposition is also beneficial for explaining the observed extended soft gamma-ray emission of some SGRBs. %Since these SGRBs probably originate from mergers of two NSs,
If these suspected post-merger NSs are true, then the upper limit of NS masses would be significantly higher than $\sim2M_{\odot}$, which indicates a very stiff equation of state for the post-merger NSs.

Besides contributing internal afterglow emission and energizing an SGRB external shock at the axial direction, the PW from a post-merger NS can also heat the nearly-isotropic merger ejecta with a high efficiency, because the merger ejecta is opaque at early times and can absorb the wind emission. Therefore, the thermal emission of the merger ejecta could be enhanced significantly (Yu et al. 2013; Metzger \& Piro 2014; Li \& Yu 2016), while this emission was previously predicted to be mainly powered by the radioactive decays of r-process elements (Li \& Paczy¨½ski 1998; Metzger et al. 2010). Here the r-process elements can be efficiently synthesized in the merger ejecta because of the high neutron fraction. Due to the combining effect of the different energy sources, the thermal transient emission of merger ejecta is suggested to be generally termed as ``mergernovae" (Yu et al. 2013), in order to distinguish from the terminology ``kilonova" which represents the ejecta emission only powered by radioactivity. In any case, since the parameters of specific post-merger NSs could vary in wide ranges, the luminosity of mergernovae can in principle range from $10^{41}{\rm erg~s^{-1}}$ to  $10^{44}{\rm erg~s^{-1}}$.

\begin{figure}
\centering\resizebox{0.5\hsize}{!}{\includegraphics{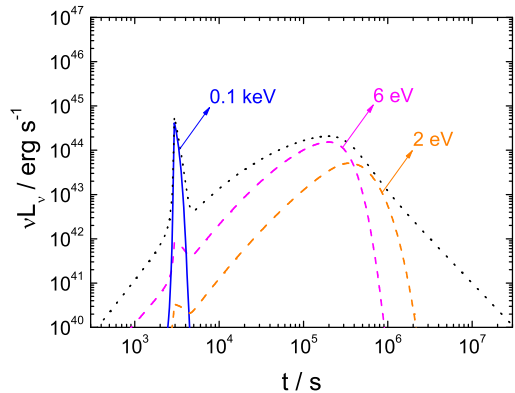}}
\caption{The bolometric (dotted) and chromatic light curves of a mergernova emission driven by the spin-down of a post-merger NS (Li \& Yu 2016).
}\label{cartoon}
\end{figure}

\begin{figure}
\centering\resizebox{0.4\hsize}{!}{\includegraphics{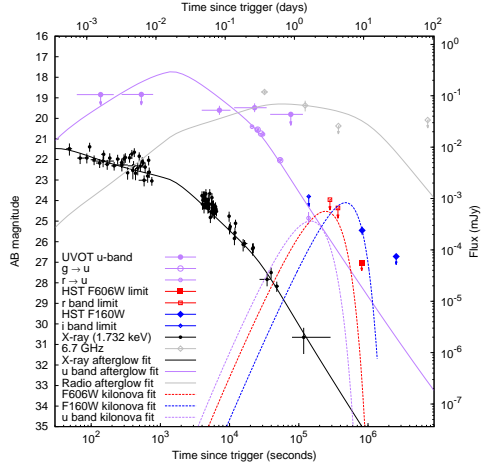}}
\caption{The multi-wavelength afterglows of GRB 130603B and the associated mergernova candidate (Fan et al. 2013).
}\label{cartoon}
\end{figure}
During the past several years, some mergernova/kilonova
candidates have been tentatively identified from the optical-infrared
emission in excess of the afterglow emission of SGRBs, e.g., GRB 130603B (Berger et al. 2013; Tanvir et al.
2013), 050709 (Jin et al. 2016), 060614 (Jin et al. 2015;
Yang et al. 2015), 050724, 061006, 070714B, and 080503 (Gao et al. 2015, 2017). For the first-discovered candidate from GRB 130603B, although it was usually regarded as a radioactivity-powered kilonova, a spin-down power is actually clearly indicated by the shallow-decay afterglow of this SGRB. By using multi-wavelength fittings, Fan et al. (2013) proved that GRB 130603B is a good example owning an NS engine, where the NS engine can be simultaneously responsible for the GRB afterglow emission and the optical/infrared excess. Furthermore, Gao et al. (2015, 2017) found that, in some SGRB afterglows, an X-ray bump can appear to associate with an optical excess emission. Here, while the optical excess is ascribed to a mergernova emission, the X-ray re-brightening can be naturally explained by the leakage of the PW emission from the merger ejecta as the ejecta becomes transparent for the X-rays.

\begin{figure}
\centering\resizebox{0.4\hsize}{!}{\includegraphics{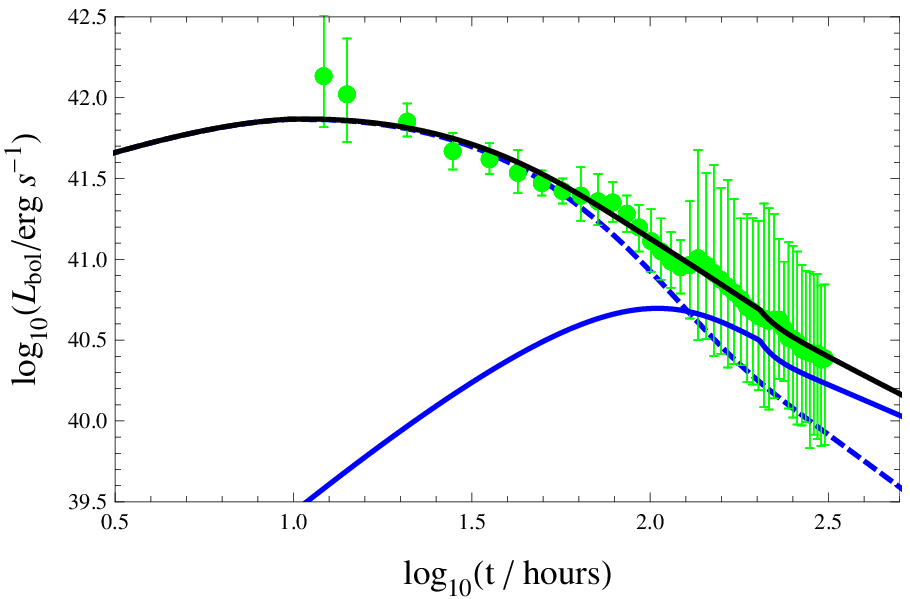}}\resizebox{0.4\hsize}{!}{\includegraphics{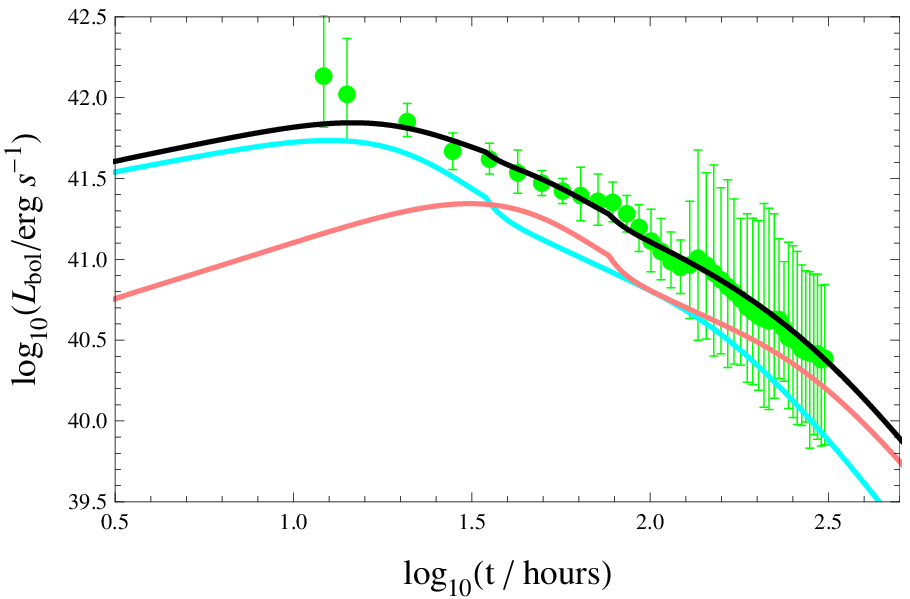}}
\caption{{\it Left:} Mergernova emission powered by the radioactivity and NS spin-down subsequently for a uniform merger ejecta, as suggested in Yu et al. (2018); {\it Right:} Mergernova emission powered by the NS spin-down only for a two-component merger ejecta, as suggested in Li et al. (2018). The data of the bolometric emission of AT 2017gfo are taken from Kasliwal et al. (2017).}\label{fig3b}
\end{figure}
The most conspicuous mergernova event is undoubtedly the optical transient AT 2017gfo (Abbott et al. 2017c; Arcavi
et al. 2017; Andreoni et al. 2017; Chornock et al. 2017; Coulter
et al. 2017; Covino et al. 2017; Cowperthwaite et al. 2017;
Drout et al. 2017; Evans et al. 2017; Hu et al.
2017; Kasliwal et al. 2017; Kilpatrick et al. 2017; Nicholl
et al. 2017; Pian et al. 2017; Shappee et al. 2017; Smartt
et al. 2017; Soares-Santos et al. 2017; Tanvir et al. 2017), which was captured by the follow-up observations after the discovery of GW170817 and GRB 170817A (Abbott et al. 2017a, b; Goldstein
et al. 2017; Savchenko et al. 2017; Zhang et al. 2017). Different from the previous mergernova candidates, AT 2017gfo was observed timely, lastingly, deeply, and in multi-wavelengths. So, this event provides an unprecedented opportunity to
probe the details of mergernova emission. On the one hand, AT 2017gfo was widely classified to a purely radioactivity-powered mergernova, i.e., a kilonova. Specifically, its emission can be modeled by invoking an ejecta mass of $\sim0.065M_{\odot}$ and a two-component ejecta of two different opacities, although this ejecta mass could be too high to be produced by a NS merger and the opacities also need to be fine tuned. In this case, the merger product of this event is suggested to be a black hole, at least, after a few seconds from the merger. On the other hand, however, the existence of a post-merger NS has not been ruled out. The invoking of a spin-down power can actually help to effectively overcome the difficulties of the kilonova model. To be specific, the energy supply of the AT 2017gfo emission could be dominated by the radioactivity and the NS spin-down subsequently (Yu et al. 2017). The spin-down-dominated emission was delayed just because all of the spin-down energy should be injected and diffused from the bottom of the merger ejecta. In this case, a complicated structure may not be necessary for the emitting ejecta. As another possibility, we suggested that the AT 2017gfo emission could be always powered by the NS spin-down. In this case, while the observations can still be well modeled, the unreasonable constraint on the ejecta masses can be loosed completely (Li et al. 2018). In summary, in our opinion, the AT 2017gfo emission somewhat provided an evidence for the formation of a long-lived massive NS from the GW170817/GRB 170817A event. The mass of the NS is around $\sim2.6M_{\odot}$.

%Anyway, in order to solve this merger product problem, more kilonova samples are needed. Personally, in my opinion, AT2017gfo is first-discovered kilonova and it must not be the most luminous kilonova in the universe. So, for more luminous kilonovae, a pure radioactive energy source will be very difficult. Seriously, in the future, on-axis observations and observations of the increasing phase of kilonova will be helpful for constraining the properties of the energy sources.

GRB 170817A could be not intrinsically different from typical SGRBs and its low luminosity could be just resulted from an off-axis observation. If this is true, then the remnant massive NS is very likely to be a magnetar intrinsically. However, the luminosity of AT 2017gfo requires that the spin-down power of the massive NS cannot be very high, at least, on the timescale of a few days. This conflict may indicate that the MDR of the NS had been seriously suppressed at the timescale of a few days, although the MDR could be very powerful at early times. As a possibility, the switch-off of the MDR could be caused by a fall-back accretion onto the NS by burying and expelling the opening field lines. Additionally, it worths to be mentioned that, due to the ultrahigh internal magnetic field, the global deformation of the NS could be significant, which can lead to an effective secular GW radiation. Therefore, in a word, before the switch-off of the MDR, it can dominate the spin-down of the NS and the energy would be consumed to accelerate the GRB jet and the merger ejecta. After this stage, the spin-down would be controlled by the GW radiation and the merger ejecta can only be heated by a significantly-suppressed PW emission.

\begin{figure}
\centering\resizebox{0.7\hsize}{!}{\includegraphics{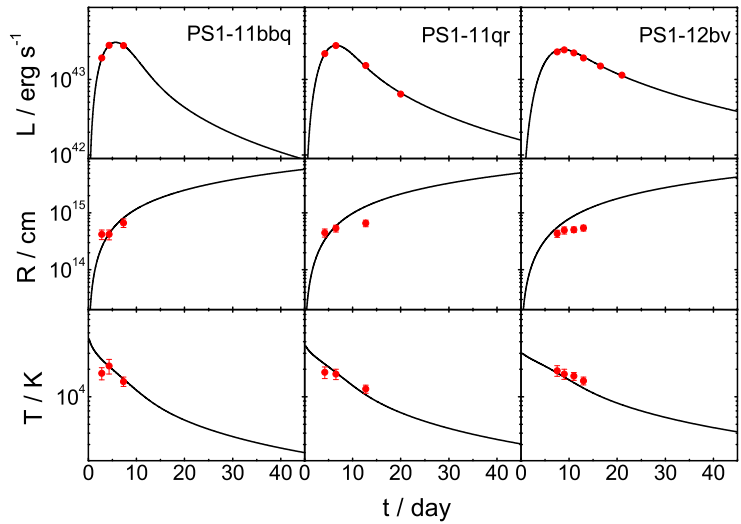}}\caption{Fitting to the light curves of three FLTs discovered in PS1-MDS by using a NS spin-down power and a low-mass ejecta (Yu et al. 2015).
}\label{fitting}
\end{figure}

\begin{figure}
\centering\resizebox{0.5\hsize}{!}{\includegraphics{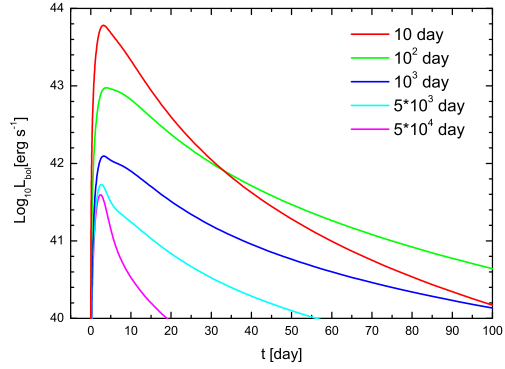}}
\caption{Bolometric light curves of AIC optical transients for different spin-down timescales of the newborn NS (Yu et al. 2019a).
%From top to bottom, the spin-down timescale is taken to be 10, 100, 1000, 5000, and $5\times10^4$ days, respectively.
}\label{Lbol}
\end{figure}

\section{Fast Evolving Luminous Optical Transients}\label{sec4}
In contrast to the beamed SGRB emission, in principle, we can have much more chances to detect orphan mergernovae unassociated by a SGRB. Interestingly, some mysterious FLTs have been indeed discovered in the past decade by some high-cadence transient surveys (Drout et al. 2014; Vink{\'o} et al. 2015; Pursiainen et al. 2018; Rest et al.
2018). The high luminosities and the short timescales of these FLTs make them difficult to be accounted for by a radioactive power, as the mass of the emission material cannot be very much higher than $0.1M_{\odot}$. Therefore, as considered for SLSNe, the invoking of a newborn NS will be very helpful for understanding these FLTs (Yu et al. 2015), while the CSM-interaction model has also been usually employed (Chevalier \& Irwin
2011; Balberg \& Loeb 2011; Ginzburg \& Balberg 2014). In any case, being inferred by the timescales of FLTs much shorter than those of SLSNe, the low masses of FLT ejecta still make them completely different from SLSNe and probably be associated with a compact object origin. Return to the discussion at the beginning of this section, some observed FLTs could in principle originate from mergers of double NSs, as candidates of orphan mergernova emission. This hypothesis can be somewhat tested by searching the off-axis GRB afterglows from these FLTs.

\section{Accretion-induced Collapses of White Dwarfs }\label{sec4}
Fairly speaking, the double NS mergers are not the exclusive origin of a system consisting of a newborn NS and a low-mass ejecta. For example, a very competitive model is the accretion-induced collapses (AICs) of white dwarfs (WDs.)
A WD in a binary can increase its mass toward
the Chandrasekhar limit by accreting material from
the companion. Besides the usual Type Ia supernova channel, the WD could
sometimes collapse into a NS by losing its
pressure support abruptly, if the electron captures in the WD core can take place more quickly
than the nuclear burning (Canal \& Schatzman 1976;
Miyaji et al. 1980; Canal et al. 1990; Nomoto \& Kondo
1991; Wang 2018). Accompanying with the formation of the NS, a small mass of $\sim10^{-3}-10^{-2}M_{\odot}$ can be ejected by a core bounce on the
proto-NS (Woosley \& Baron 1992; Dessart et al. 2006)
and sometimes by a wind from the accretion
disk (Metzger et al.
2009). In analogy to the NS merger cases, this low-mass AIC ejecta can be effectively heated by the NS spin-down and the radioactive decays of $^{56}$Ni (Yu et al. 2019a,b). Therefore, FLT emission can be naturally produced by the AIC ejecta.

\section{Summary}\label{sec4}
NSs in the universe could have very different origins, including core-collapse supernovae, NS mergers, WD collapses, and the so-called electron-capture supernovae that are somewhat similar to the AICs, etc. Accompanying with the NS formations, an explosive material ejection can always take place and some radioactive elements can be synthesized. The decays of the radioactive elements can always heat the ejecta. Furthermore, the ejecta can also be energized by the spin-down of the newborn NSs, probably more effectively if the NSs can rotate very quickly. As a result, some unusual optical transient phenomena including SLSNe and FLTs can be produced by the ejecta emission. These transients are different from each other because of the different NS parameters and ejecta masses. For the FLTs, many progenitor models have been proposed including NS mergers and WD AICs etc. The distinguishing of these competitive models requires more observational samples and multi-wavelength counterpart observations (Yu et al. 2019b). In extreme situations, if the dipolar magnetic fields of the newborn NSs can arrive at $\sim10^{15}$G, then a relativistic jet could be launched. If the jet can successfully escape from the surrounding material, then an GRB can be produced to be associated with the unusual optical transients. In summary, observations of these different transients provide an available and valuable route to explore the nature of newborn NSs.

\section{Acknowledgement}
The authors appreciate Bing Zhang for his collaboration on some topics. This work is
supported by the National Natural Science Foundation
of China (grant No. 11473008 and 11573014) and the
self-determined research funds of CCNU from the colleges'
basic research and operation of MOE of China.

% References

%\nocite{*}
%\bibliographystyle{aipnum-cp}%
%\bibliography{sample}%

\end{document}